\documentclass[fleqn, usenatbib]{mnras}

\usepackage{newtxtext}

\usepackage[T1]{fontenc}
\usepackage{ae,aecompl}

\usepackage{graphicx}	
\usepackage{amsmath}	
\usepackage{amssymb}	
\usepackage{subfig}
\usepackage[font=footnotesize]{caption}
\usepackage{graphicx}
\usepackage{wrapfig}
\usepackage{gensymb}
\usepackage{comment}
\usepackage{multirow}
\usepackage{tabularx}
\usepackage{footnote}
\usepackage[dvipsnames]{xcolor}
\usepackage{stackengine}
\usepackage{multirow}
\usepackage{tablefootnote}
\usepackage{enumitem}
\usepackage{longtable}
\usepackage{threeparttablex}
\usepackage{hhline}
\usepackage{bm}		
\usepackage{pdflscape}	

\def\maxithirt{MAXI~J1348--630}
\def\maxieight{MAXI~J1820$+$070}

\def\hh{H1743--322}
\def\gx{GX~339--4}

\title[\maxithirt{} in quiescence]{The black hole X-ray binary \maxithirt{} in quiescence}

\author[F. Carotenuto et al.]{F. Carotenuto,$^{1, 2}$\thanks{E-mail: francesco.carotenuto@physics.ox.ac.uk}
S. Corbel,$^{2, 3}$
A. Tzioumis$^{4}$
\\
$^{1}$Astrophysics, Department of Physics, University of Oxford, Keble Road, Oxford OX1 3RH, UK\\
$^{2}$Universit\'{e} Paris Cit\'{e} and Universit\'{e} Paris Saclay, CEA, CNRS, AIM, F-91190 Gif-sur-Yvette, France\\
$^{3}$ORN, Observatoire de Paris, Université PSL, Université Orléans, CNRS, 18330 Nançay, France\\
$^{4}$Australia Telescope National Facility, CSIRO, PO Box 76, Epping, New South Wales 1710, Australia\\
}
\date{Accepted XXX. Received YYY; in original form ZZZ}

\pubyear{2022}

\begin{document}
\label{firstpage}
\pagerange{\pageref{firstpage}--\pageref{lastpage}}
\maketitle

\begin{abstract}
\noindent The properties of the disk/jet coupling in quiescent black hole low mass X-ray binaries (BH LMXBs) are still largely unknown.
In this paper we present the first quasi-simultaneous radio and X-ray detection in quiescence of the BH LMXB \maxithirt{}, which is known to display a \textit{hybrid} disk/jet connection that depends on the accretion rate. We performed deep X-ray and radio observations using the \textit{Chandra} X-ray Observatory and the Australia Telescope Compact Array. \maxithirt{} is detected for the first time in quiescence at an X-ray luminosity $L_{\rm X} = (7.5 \pm 2.9) \times 10^{30} (D/2.2 \ {\rm kpc})^2$ erg s$^{-1}$: one of the lowest X-ray luminosities observed for a quiescent BH LMXB, possibly implying a short orbital period for the system. \maxithirt{} is also detected in radio at $L_{\rm R} = (4.3 \pm 0.9) \times 10^{26} (D/2.2 \ {\rm kpc})^2$ erg s$^{-1}$. These detections allow us to constrain the location of \maxithirt{} on the radio/X-ray diagram in quiescence, finding that the source belongs to the standard (\textit{radio-loud}) track in this phase. This provides a strong confirmation that hybrid-correlation sources follow the standard track at low luminosities and down to quiescence, thus improving our knowledge of the disk/jet connection in BH LMXBs.
\end{abstract}


\section{Introduction}
\label{sec:Introduction}

Black holes (BH) low mass X-ray binaries (LMXBs) are generally observed when they enter into bright outbursts. During such phases, the X-ray luminosity from the accretion disk increases by several orders of magnitude, and the systems transition between different accretion states with well-defined spectral and timing properties \citep{Homan_belloni, Remillard_xrb, Tetarenko_2016, Belloni_Motta_2016}. In addition, synchrotron-emitting compact jets and moving plasma bubbles, which carry away a significant fraction of the accretion power, are observed in the radio band during, respectively, the hard state and the hard-to-soft state transition (e.g.\ \citealt{Corbel_2000, Fender_2001, Fender2006}). However, BH LMXBs are for most of their time in a quiescent state, with low accretion rates and X-ray luminosities in the range between $10^{30}$ and $10^{34}$ erg s$^{-1}$, corresponding to Eddington fractions of $10^{-9} \sim 10^{-5}$ for a 10 $M_{\odot}$ BH (e.g.\ \citealt{Plotkin2013, Reynolds_2014, Plotkin_2021}). While quiescence is the most common state for a BH LMXB, the properties of the accretion flow and of the jets at low luminosities still need to be properly characterized.

When in quiescence, the infrared and optical emission from these systems is dominated by the companion star, while compact jets, albeit faint, are still present and can be detected in the radio band (e.g.\ \citealt{Plotkin_2017_v404, Gallo2019}). It appears that the jet properties are essentially the same between outburst and quiescence, with the radio spectrum that might change from flat-to-inverted to moderately steep \citep{Corbel2013_corr, Tremou2020}. This supports the current picture of the quiescence state as a sort of low-luminosity version of the hard state, with very similar spectral and physical properties (e.g.\ \citealt{Gallo2019}). Moreover, the compact jets in quiescence might channel a larger fraction (with respect to the outburst phase) of the radiative power of the accretion flow, strongly motivating radio observations of quiescent accreting compact objects (e.g.\ \citealt{Plotkin_2019}). The dominance of the jets at these low luminosities is also believed to be responsible for an observed progressive X-ray spectral softening (e.g.\ \citealt{Corbel2006, Plotkin2013}). In analogy with the hard state, the properties of the accretion flow should also be similar, and, according to one of the main proposed scenarios, the BH could be fed by a hot radiatively inefficient accretion flow (e.g.\ \citealt{Narayan_1994, Esin}).

Compact jets from BH LMXBs are fundamentally connected to the accretion flow, as the levels of jet emission strongly depends on the accretion rate, both in outburst and in quiescence. This is observed in the hard state as a non-linear correlation between the radio and X-ray luminosities taking the form of a power law ($L_{\rm R} \propto L_{\rm X}^{\beta}$). Such correlation is observed over several orders of magnitude in luminosity on the so-called radio/X-ray diagram (e.g.\ \citealt{Hannikainen_1998, Corbel_2000, Corbel_2003}), and it is today an important diagnostic of the accretion/ejection coupling in BH LMXBs, which can also be used to constrain the emission mechanisms in these sources. Two main groups of BHs are currently observed on the radio/X-ray diagram, displaying two distinct correlations with different power law indices $\beta$. The upper track, historically called \textit{standard}, but often referred to as \textit{radio-loud}, is characterised by $\beta \simeq 0.6$ and includes well-known sources as \gx{}, V404~Cyg and \maxieight{} \citep{Corbel_2000, Corbel_2003, Corbel2013_corr, Gallo_2003, Bright, Shaw_2021}. On the other hand, sources on the lower track are labeled as \textit{outliers} (or \textit{radio-quiet}) and display indices $\beta \simeq 1-1.4$ \citep{Coriat, Jonker_MAXI, Gallo_2014, Monageng_2021, Carotenuto_lrlx}. The majority of BH LMXBs appears now to belong to this second group. Interestingly, thanks to dense monitoring campaigns, some of the outliers were found to display a \textit{hybrid} correlation, transitioning between the two tracks at different X-ray luminosities, between $10^{-4}$ and $10^{-3} L_{\rm Edd}$ \citep{Coriat, Carotenuto_lrlx}. While the size of the available data sets of BH LMXBs on the radio/X-ray diagram increased considerably in the recent years, the reasons for the observed distribution of sources on the diagram are still unclear. Possible explanations include geometric effects, different jet physical properties and different accretion flow properties (e.g.\ \citealt{Peer_Casella_2009, Coriat, Xie_2016, Motta_2018}).

Much less is known about the properties of the two groups of sources in quiescence, since the quiescent regions of the radio/X-ray diagram are still greatly unexplored, due to the extreme faintness of quiescent BH LMXBs. So far, only a handful of BH LMXBs have been detected in quiescence, mostly appearing to be compatible with the standard track (e.g.\ \citealt{Gallo_2006, Miller-Jones_2011, Corbel2013_corr, Ribo_2017, Tremou2020}), while the behaviour of outliers is completely uncertain. Outlier (or hybrid) sources  could re-join the standard track in quiescence \citep{Coriat}, but this has still to be confirmed with additional observations.

In this Letter we present the first radio and X-ray detection in quiescence of \maxithirt{}, constraining its location on the diagram at the lowest possible luminosities. \maxithirt{} is a BH LMXB discovered in 2019, when it entered into a bright outburst \citep{Tominaga_1348}. The source appears to be located at a close distance, between 2.2 and 3.4 kpc \citep{Chauhan2021, Lamer_2021}, which is favourable for a study of the system in quiescence. In \cite{Carotenuto2021} we have presented the full radio and X-ray monitoring of \maxithirt{} during its 2019/2020 outburst, and we refer to that paper for details on the various outburst phases. Our radio observations detected and covered the entire evolution of compact jets in the outburst phase, during which \maxithirt{} also displayed some extremely energetic large-scale discrete ejecta \citep{Carotenuto2021, Carotenuto_2022}. Combining the detections of compact jets with the quasi-simultaneous X-ray observations, we were able to cover the whole evolution of \maxithirt{} on the radio/X-ray diagram, finding that this source belongs to the numerous group of outliers. In fact, \maxithirt{} displays a clear hybrid radio/X-ray correlation, re-joining the standard track at $L_{\rm X} \simeq 10^{33}$ erg s$^{-1}$, and it is today the hybrid-correlation source with the most detailed coverage on the diagram, and such coverage was obtained during a single outburst  \citep{Carotenuto_lrlx}. Therefore, constraining its behaviour in quiescence is critical for completing the full track of the source on the radio/X-ray diagram, which is something unprecedented an hybrid source.

\section{Observations}
\label{sec:Observations}

To constrain the quiescent level of \maxithirt{} in radio and X-rays, we performed a quasi-simultaneous observation of the source with the NASA \textit{Chandra} X-ray observatory \citep{Weisskopf_2000} and with the  Australia Telescope Compact Array (ATCA, \citealt{Frater_1992}).

\subsection{\textit{Chandra} observation}
\label{sec:Chandra_observation}

The \textit{Chandra} observations of \maxithirt{} were performed with the Advanced CCD Imaging Spectrometer (ACIS-S) on 2021 June 28 (30 ks; ObsId 25054) under Director's Discretionary Time (PI: F.\ Carotenuto). We restricted the observation to the back-illuminated chip S3, providing the best low-energy response. The X-ray data analysis was performed using the \textit{Chandra} Interactive Analysis of Observation (CIAO) software 4.14.1 \citep{Fruscione2006}, with the calibration files CALDB version 4.9.6. We filtered to only keep the events in the energy range 0.3--8 keV. No background flare was detected. The \texttt{fluximage} script was used to create the X-ray images, keeping the bin size to 1 (1 pixel = 0.492\arcsec) (see Fig. 1). A single X-ray source is detected at a position consistent with \maxithirt{}.  

We extracted an energy spectrum in the 0.3--8 keV energy range from a circular source region with a 2\arcsec \ radius, while the background was extracted from an annulus with an inner radius of 10\arcsec \ and an outer radius of 20\arcsec. Spectral analysis was carried out with XSPEC \citep{Arnaud_xspec}. Our main goal with the X-ray spectral analysis was to estimate the unabsorbed flux in the 1--10 keV band. Hence, we fitted the spectrum with a simple absorbed power law model ({\tt tbabs$\times$powerlaw}), for which the {\tt tbabs} component accounts for interstellar absorption, modelled with an equivalent hydrogen column density ($N_{\rm H}$) by using {\tt wilm} abundances \citep{Wilms} and {\tt vern} cross-sections \citep{Verner1996}. Concerning the spectral parameters, the power law photon index was fixed to $\Gamma = 2.2$, in accordance with what is generally measured for BH LMXBs in quiescence \citep{Corbel2006, Corbel2008, Plotkin2013}, and the column density was fixed to $ N_{\rm H} = 0.86 \times 10^{22}$ cm$^{-2}$, measured from X-ray observations during the outburst \citep{Carotenuto2021, Carotenuto_lrlx}. As generally prescribed for low-counts observations, Cash statistics {\tt cstat} \citep{cstat} was used in the fitting process. We then calculated the 1--10 keV X-ray unabsorbed flux using the XSPEC convolution model {\tt cflux}.

\begin{figure}
\begin{center}
\includegraphics[width=\columnwidth]{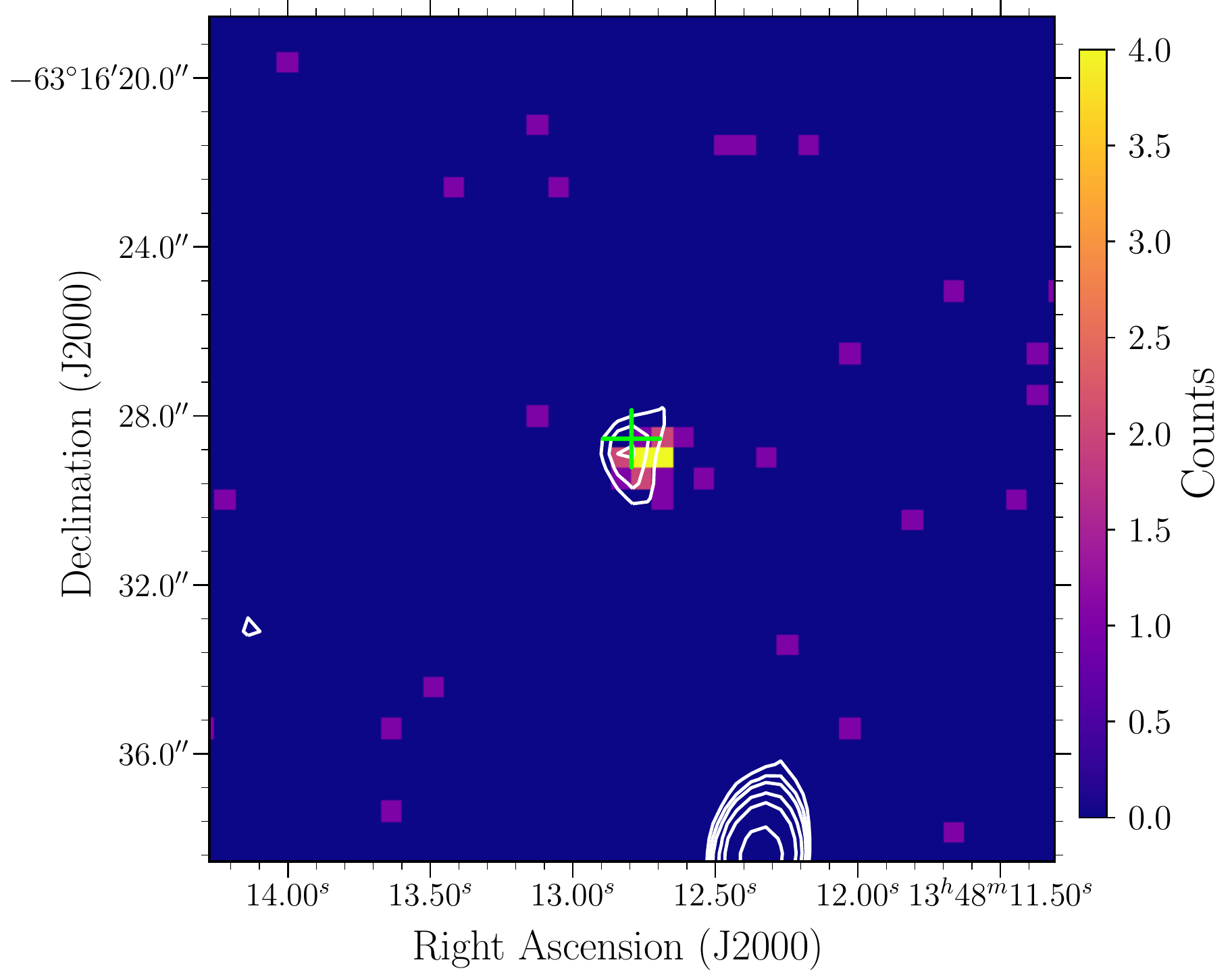}
\caption{\textit{Chandra} ACIS-S3 X-ray detection of \maxithirt{} in quiescence, in the 0.3--8 keV energy range, with overlaid ATCA radio contours. The green cross is centered on the location of \maxithirt{} \citep{Carotenuto2021}, and the units on the color bar are photon counts. The radio image is obtained by stacking the simultaneous observations in C and X bands, with 24 hours of exposure time on source. Contours start a 3 times the rms (3 $\mu$Jy beam$^{-1}$). The source is detected in radio with a significance of 5$\sigma$.}
\label{fig:quiescence_detection}
\end{center}
\end{figure}

\subsection{ATCA observation}
\label{sec:ATCA observation}

A long ATCA observation was performed between 3 and 5 July 2021, less than a week after the \textit{Chandra} observation, for a total of 24 hours on source (proposal ID C3416, PI Carotenuto). During this interval, the array was in the 6B extended configuration.
The observation was taken simultaneously in the C and X bands, at central frequencies of 5.5 and 9 GHz, respectively. For each central frequency, the total bandwidth was 2 GHz. PKS~1934--638 was used for bandpass and flux calibration, while PKS~1352--63 was used for the complex gain calibration. The data were first flagged and then calibrated using standard procedures with the Common Astronomy Software Application (\textsc{casa}, \citealt{CASA}). Imaging was carried out with the standard {\tt tclean} algorithm in \textsc{casa} with a natural weighting scheme chosen in order to maximize sensitivity. 
To obtain the lowest possible rms, we stacked the two bands into a single image, adopting the Multi-term (Multi Scale) Multi-Frequency Synthesis (MTMFS, or MS-MFS, \citealt{Rau_2011}) as a deconvolution algorithm in order to take into account the large fractional bandwidth, using two Taylor terms. In this configuration, the rms reached 3 $\mu$Jy beam$^{-1}$. To obtain the radio flux density $S_{\nu}$, we fitted a point source in the image plane with the \textsc{casa} task {\tt imfit}.

\section{The quiescent level of \maxithirt{}}
\label{sec:The quiescent level of MAXIJ1348}

\begin{figure*}
\begin{center}
\includegraphics[width=0.85\textwidth]{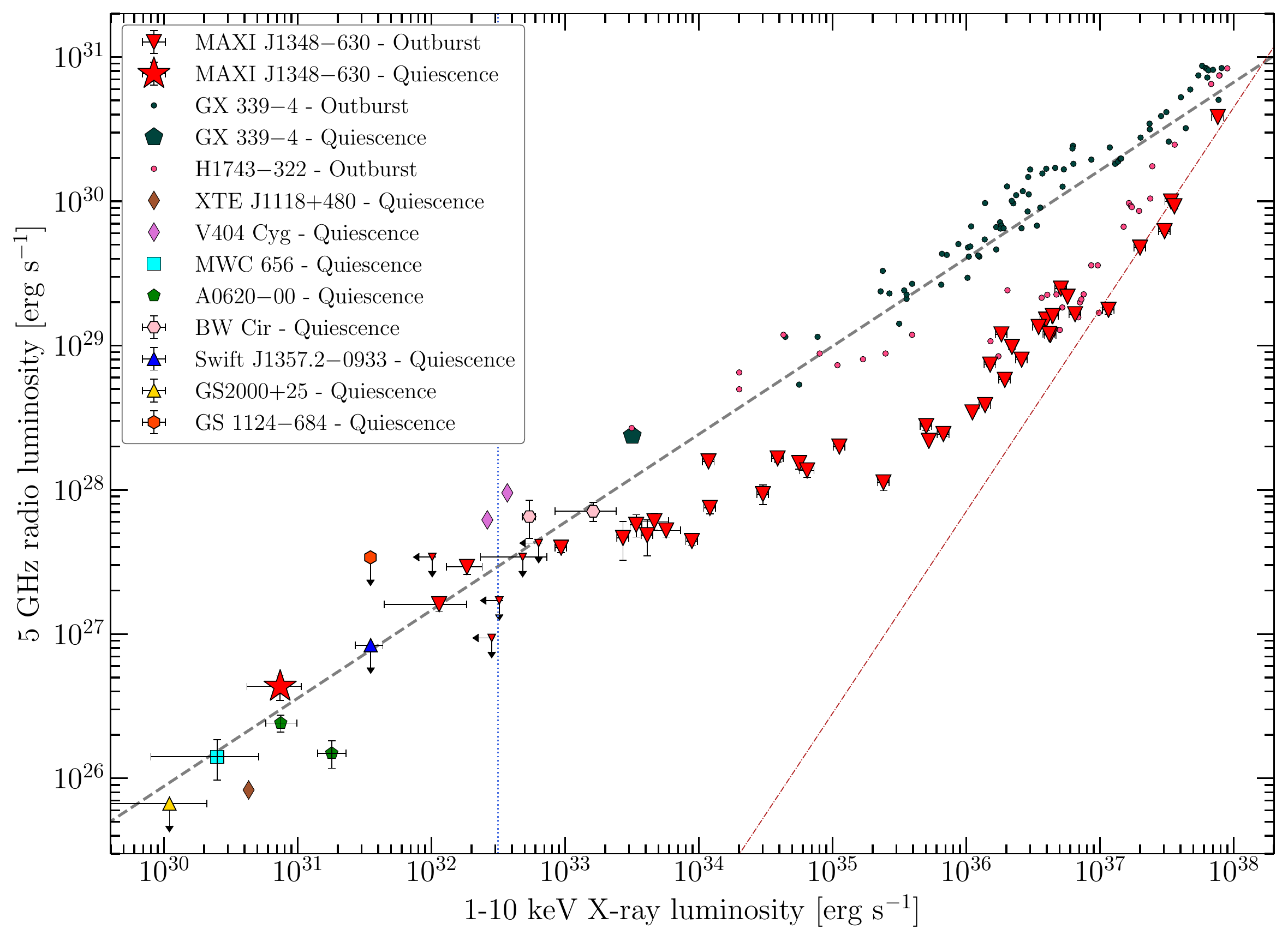}
\caption{Radio/X-ray diagram displaying \maxithirt{} during its 2019/2020 outburst and in quiescence. For consistency with \citet{Carotenuto_lrlx}, we assume $D = 2.2$ kpc \citep{Chauhan2021}. The vertical dotted line marks the inferred transition luminosity to the standard track (albeit with a large uncertainty, \protect\citealt{Carotenuto_lrlx}), while the red and grey dashed lines represent, respectively, the standard ($\beta = 0.6$) and outlier ($\beta = 1.4$) tracks, which are shown for illustrative purposes. \maxithirt{} is on the standard track in quiescence. For comparison, \hh{} \citep{Jonker_2010, Coriat} and \gx{} \citep{Corbel2013_corr} are shown on the diagram with smaller size points. Sources in quiescence are marked with different colors. The plot shows the location on the diagram in quiescence for A0620--00 \citep{Gallo_2006}, MWC~656 \citep{Ribo_2017}, XTE~J1118+480 \citep{Gallo_2014}, \gx{} \citep{Tremou2020}, BW~Cir \citep{Plotkin_2021} and V404~Cyg \citep{Plotkin_2017_v404}, as well as upper limits for GS~2000+25 \citep{Rodriguez_2020}, GS~1124--684 \citep{Plotkin_2021} and Swift J1357.2-0933 \citep{Plotkin_2016, Charles_2019}.}
\label{fig:x_radio_corr_quiescence}
\end{center}
\end{figure*}

We obtained a significant detection of \maxithirt{} in quiescence, both in radio and X-rays. The detection is shown in Figure \ref{fig:quiescence_detection}. In the X-rays, 21 counts from \maxithirt{} were collected by \textit{Chandra} (0.3--8 keV) in the 30 ks of exposure, implying an unambiguous detection at the lowest level of activity of the source. After spectral fitting (Section \ref{sec:Chandra_observation}), we obtain an integrated unabsorbed flux $F_{\rm X} = (1.3 \pm 0.5) \times10^{-14}$ erg cm$^{-2}$ s$^{-1}$ in the 1$-$10 keV energy range. This can be then converted to the integrated luminosity $L_{\rm X} = 4\pi D^2  S_{\rm X }$. For consistency with \cite{Carotenuto_lrlx}, in this paper we assume distance $D = 2.2$ kpc \citep{Chauhan2021} and we quote the luminosities with a factor $(D/2.2 \ {\rm kpc})^2$ in order to include the distance estimation by \cite{Lamer_2021}. We obtain $L_{\rm X} = (7.5 \pm 2.9) \times 10^{30} (D/2.2 \ {\rm kpc})^2$ erg s$^{-1}$.

\maxithirt{} has also been detected in radio during its quiescence phase. As can be seen from Figure \ref{fig:quiescence_detection}, with 24 h of ATCA exposure time, \maxithirt{} is detected as a point source at its known location, at a significance level of $5\sigma$ and with a peak flux of $15.3 \pm 2.5$ $\mu$Jy. The faint radio emission coming from the core location can be safely attributed to the presence of compact jets in quiescence, as observed for other BH LMXBs \citep{Corbel2013_corr, Gallo2019, Tremou2020}. We then converted the measured radio flux density $S_{\nu}$ to the 5 GHz monochromatic luminosity $L_{\rm R} = 4\pi D^2  \nu S_{\nu}$, conservatively assuming a flat radio spectrum. This leads to a monochromatic 5 GHz radio luminosity $L_{\rm R} = (4.3 \pm 0.9) \times 10^{26} (D/2.2 \ {\rm kpc})^2$ erg s$^{-1}$.

With the quasi-simultaneous radio and X-ray detection in quiescence, we obtain a new, critical measurement for \maxithirt{}, completing the coverage of this source on the radio/X-ray diagram. The updated radio/X-ray diagram is shown in Figure \ref{fig:x_radio_corr_quiescence}. One can immediately notice that \maxithirt{} aligns very well with the standard track defined by \gx{}, V404~Cyg and \maxieight{} (e.g.\ \citealt{Corbel2013_corr}). Along the detection in quiescence, the clear confirmation that \maxithirt{} lies on the standard track in quiescence is one of the main results of this work.
We performed a a simple power law fit $L_{\rm R} \propto L_{\rm X}^{\beta}$ to the three points on the diagram with $L_{\rm X} < L_{\rm stand} = 3.2^{+61.3}_{-3.0} \times 10^{32} (D/2.2 \ \ {\rm kpc})^2$ erg s$^{-1}$ (see Figure \ref{fig:x_radio_corr_quiescence} and \citealt{Carotenuto_lrlx}), including the detection in quiescence and the two lowest detection in the outburst lying on the standard track. The fit, relying on {\tt curve$\_$fit} from the SciPy package, yields $\beta = 0.57 \pm 0.08$, which is fully consistent with the standard track (e.g.\ \citealt{Corbel2013_corr}).

\section{Discussion}
\label{Discussion}

\maxithirt{} is one of the few BH LMXBs detected in quiescence both in the radio and X-ray bands. Due to the need of quasi-simultaneous deep observations, these detections are exceedingly rare at present. Even if a model-independent mass estimation of the BH is not yet available, the quiescence X-ray luminosity of \maxithirt{} corresponds to an Eddington fraction $\sim$10$^{-8.3} L_{\rm Edd}$, assuming a $10 M_{\odot}$ BH. Currently, the BH LMXB with the lowest known X-ray luminosity in quiescence is GS~2000+25, detected with \textit{Chandra} at $\sim$10$^{30}$ erg s$^{-1}$, namely $\sim$10$^{-9} L_{\rm Edd}$ \citep{Rodriguez_2020}. The detections at such low Eddington fractions are among the lowest level of activity ever explored for an accreting compact object. However, for GS~2000+25 a radio counterpart was not detected \citep{Rodriguez_2020}.

The most interesting result of this work revolves around the location of \maxithirt{} on the radio/X-ray diagram in quiescence. In \cite{Carotenuto_lrlx}, we demonstrated that \maxithirt{} belongs to the less-explored group of the so-called \textit{hybrid-correlation} sources. With this detection, we can confirm for the first time that these sources follow the standard track down to quiescence. Moreover, after this work and the coverage presented in \cite{Carotenuto_lrlx}, \maxithirt{} becomes the first BH LMXB to be characterised on the radio/X-ray diagram for over seven orders of magnitude in X-ray luminosity and almost four orders of magnitude in radio luminosity, during a single outburst and in quiescence. The continuity of evolution on the standard track from the end of the outburst (see Figure \ref{fig:x_radio_corr_quiescence} and \citealt{Carotenuto_lrlx}), and the displayed power law X-ray spectrum, are consistent with the current picture of the quiescence state as a low-luminosity analogue of the hard state (e.g.\ \citealt{Gallo2019}). 

It is then possible to compare the location of \maxithirt{} on the diagram with the handful of other BH LMXBs for which radio and X-ray detections in quiescence are available, shown in Figure \ref{fig:x_radio_corr_quiescence}. \maxithirt{} appears to have a quiescence luminosity that is comparable in radio and X-rays with A0620--00 \citep{Gallo_2006}, MWC~656 \citep{Ribo_2017} and XTE~J1118+480 \citep{Gallo_2014}. Stringent radio upper limits are also available for Swift~J1357.2--0933 \citep{Plotkin_2016} and GS~2000+25 \citep{Rodriguez_2020}.
On the other hand, \maxithirt{} displays a quiescent luminosity which is at least one order of magnitude lower than other very relevant quiescent BH LMXBs on the diagram, such as \gx{} \citep{Corbel2013_corr, Tremou2020} and V404~Cyg \citep{Plotkin_2017_v404}.

\begin{figure}
\begin{center}
\includegraphics[width=\columnwidth]{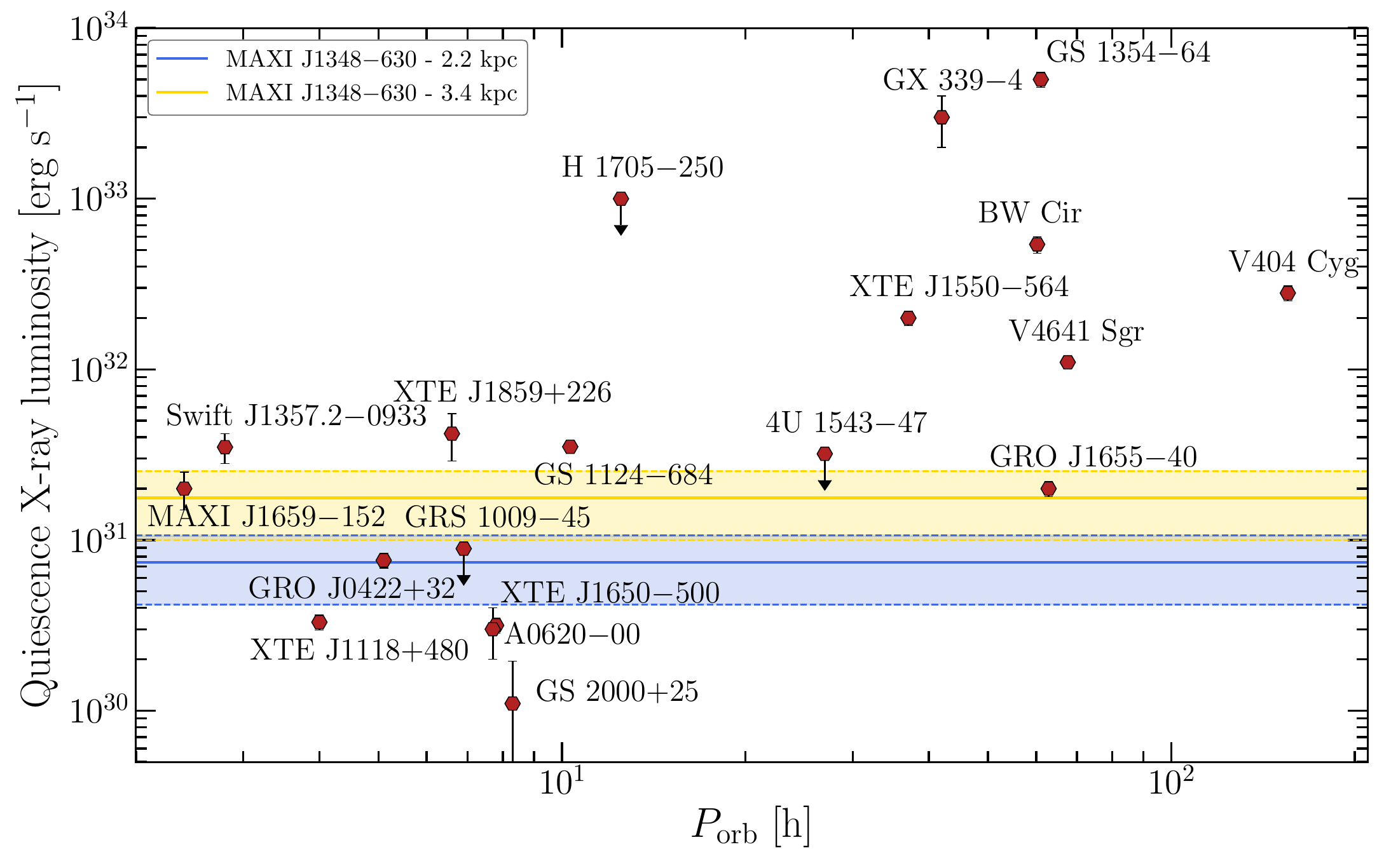}
\caption{Quiescent X-ray luminosity for \maxithirt{} (at both possible distances) compared with the population of BH LMXBs detected in quiescence with known $P_{\rm orb}$, appearing to favour a shorter $P_{\rm orb}$ for \maxithirt{}. Data from \protect\cite{Garcia_2001, Corbel2006, Reynolds_2011, Plotkin_2017_v404, Rodriguez_2020, Plotkin_2021} and references therein.}
\label{fig:p_orb_quiescence}
\end{center}
\end{figure}

A relation between the quiescent X-ray luminosity $L_{\rm X}$ and the orbital period $P_{\rm orb} $ exists for X-ray binaries \citep{Menou_1999, Garcia_2001}, where a higher $L_{\rm X}$ in quiescence is expected for higher $P_{\rm orb}$, due to the larger size of the accretion disk.
While the orbital period of \maxithirt{} is currently unknown, from its quiescence X-ray luminosity it is possible to speculate that its $P_{\rm orb}$ is likely shorter than $10 \sim 20$ h, as can be seen from Figure \ref{fig:p_orb_quiescence}, at both possible distances. This is also in line with the peak luminosity in outburst ($\sim$ $10^{-1} L_{\rm Edd}$, \citealt{Carotenuto2021}), which might suggest a similar order of magnitude for $P_{\rm orb}$ \citep{Wu_2010}.

A detection of \maxithirt{} in quiescence on the standard track also supports the findings obtained for V404~Cyg and \maxieight{} \citep{Plotkin_2017_v404, Shaw_2021}, ruling out the possibility that the X-ray emission in quiescence is dominated by the synchrotron radiation produced by the jet. Such scenario, in fact, would result in a steepening of the observed correlation at low luminosities  \citep{yuan_cui_2005}, which is instead not observed.

The X-ray spectrum of \maxithirt{} in quiescence can be adequately described with an absorbed power law with $\Gamma = 2.2$. The spectral softening when approaching quiescence is a well-known phenomenon \citep{Corbel2006, Corbel2008, Plotkin2013, Liu_index}, observed also in \maxithirt{} \citep{Carotenuto2021}. Such softening could be explained by a lack of hard X-ray photons resulting from the decrease of the inverse Compton scattering efficiency at low luminosities \citep{Esin, Qiao_2013, Liu_index}, or by a dominance at low accretion rate of Comptonization from thermal particles at the base of the jet, and not in the corona \citep{Markoff_corona, Poutanen_2014, Plotkin2015}.

Thanks to this new detection in quiescence, we obtained the complete coverage of the evolution of \maxithirt{} on the radio/X-ray diagram, from the quiescent state to the peak luminosity during the outburst. We believe that the full data set of \maxithirt{} on the diagram will turn out to be particularly important for testing different models describing the disk/jet connection in accreting BHs, leading to major improvements in our understanding of hybrid-correlation sources and, more in general, of the behaviour of BH LMXBs on the diagram.

\section*{Acknowledgements}

We thank the anonymous referee for the careful reading of the manuscript and for the valuable comments. This research has made use of data obtained from the \textit{Chandra} X-ray Observatory and software provided by the Chandra X-ray Center (CXC) in the application package CIAO.
FC, SC and AT thank Jamie Stevens and staff from the Australia Telescope National Facility (ATNF) for scheduling the ATCA radio observations. ATCA is part of the ATNF which is funded by the Australian Government for operation as a National Facility managed by CSIRO. We acknowledge the Gomeroi people as the traditional owners of the ATCA observatory site. FC acknowledges support from the Royal Society through the Newton International Fellowship programme (NIF/R1/211296) and from the project Initiative d’Excellence (IdEx) of Universit\'{e} de Paris (ANR-18-IDEX-0001). 
We acknowledge the use of the Nan\c cay Data Center, hosted by the Nan\c cay Radio Observatory (Observatoire de Paris-PSL, CNRS, Universit\'{e} d'Orl\'{e}ans), and supported by Region Centre-Val de Loire.
This project also made use of \textsc{matplotlib} \citep{matplotlib}, \textsc{numpy} \citep{harris2020array} and Overleaf (\url{http://www.overleaf.com}).

\section*{Data availability}
The un-calibrated ATCA visibility data are publicly available at the ATNF archive at \url{https://atoa.atnf.csiro.au}. The \textit{Chandra} data are instead available from the Chandra Data Archive at \url{https://cxc.harvard.edu/cda}.

\bibliographystyle{mnras}

\bibliography{maxi1348_paper4}

\bsp	
\label{lastpage}
\end{document}